\documentstyle[12pt,epsf]{article}

\begin{document}

\begin{titlepage}

\begin{flushright}
MIT-CTP-3320\\
BUHEP-02-38
\end{flushright}
\vskip 2.5cm

\begin{center}
{\Large \bf Analysis of a Toy Model of Electron ``Splitting''}
\end{center}

\vspace{1ex}

\begin{center}
{\large B. Altschul\footnote{{\tt baltschu@mit.edu}}}

\vspace{5mm}
{\sl Department of Mathematics} \\
{\sl Massachusetts Institute of Technology} \\
{\sl Cambridge, MA, 02139-4307 USA} \\

\vspace{8mm}

{\large C. Rebbi}

\vspace{5mm}
{\sl Department of Physics} \\
{\sl Boston University} \\
{\sl Boston, MA, 02215 USA} \\

\end{center}

\vspace{2.5ex}

\medskip

\centerline {\bf Abstract}

\bigskip

We examine Maris' recent suggestion that the fission of electron-inhabited
bubbles in liquid helium may give rise to a new form of electron
fractionization. We introduce a one-dimensional toy-model---a simplified
analogue of the helium system---which may be analyzed using the
Born-Oppenheimer approximation. We find that none of the model's low-lying
energy eigenstates
have the form suggested by Maris' computations, in which the bubbles were
treated completely classically. Instead, the eigenstates
are quantum-mechanically entangled superposition states, which the classical
treatment overlooks.

\bigskip 

\end{titlepage}

\newpage

Recently, Maris~\cite{ref-maris} has suggested that the splitting of an
electron-inhabited bubble in liquid helium may result in a division of the
electron into separate pieces. Each piece would behave like only a fraction
of the original electron. This phenomenon would be entirely different from the
any of the accepted forms of fermion
fractionization~\cite{ref-jackiw1,ref-su,ref-laughlin}. Maris' interpretation
has been criticized~\cite{ref-jackiw2} as corresponding to an ordinary quantum
superposition phenomenon and not indicating any fractionization of the
electron. We shall discuss this criticism further.
 
Maris considers an electron trapped in bubble in liquid helium. The
repulsive interactions between the single electron and the electron clouds of
the helium atoms support the bubble and keep the electron confined in a
deep potential well. The electron is initially in the ground (1s) state of the
well, before being excited into the 1p state. Maris considers the
time evolution of the coupled electron-bubble system after the electron is
excited.

Maris treats the electron quantum mechanically and the bubble classically. The
electron's 1p wave function is not spherically symmetric, so it exerts
different pressures at different points on the bubble. Maris treats this
pressure gradient classically and studies its effect on the bubble's shape.
He concludes that the bubble will deform and eventually split into two
separate bubbles, each containing half of the electron's wave function.

We contend that the classical analysis does not capture the correct time
evolution of the bubble's shape.
When the entire system is treated quantum-mechanically, the electron's wave
function becomes entangled with the wave
function of the bubble. The end result is not two bubbles, each containing half
of the electron, but rather a system in an entangled
state---a superposition of two bubbles in different positions. If the position
of the electron is measured, the system will collapse into a state with only
a single bubble, surrounding the location where the electron is found.

To shed light on this problem, we
shall consider a toy model which is similar to the liquid helium system in
many important ways. Our model is one-dimensional and contains only
three interacting particles, so we may learn a great deal about the behavior
of the system using simple, analytical techniques. We shall make particular
use of the Born-Oppenheimer approximation.

The system contains two heavy objects of equal mass $M$ (the ``atoms'') and
a lighter object of mass $m$ (the ``electron''), moving on a segment of length
$L$ with periodic boundary conditions. The atoms and electron interact
through repulsive potentials. If we let $x_{1}$, $x_{2}$, and $x_{e}$ be the
positions of the atoms and the electron and set $\hbar=1$, the Hamiltonian is
\begin{eqnarray}
H & = & K+V \\
K & = & -\frac{1}{2M}\frac{\partial^{2}}{\partial x_{1}^{2}}
-\frac{1}{2M}\frac{\partial^{2}}{\partial x_{2}^{2}}
-\frac{1}{2m}\frac{\partial^{2}}{\partial x_{e}^{2}} \\
V & = & V_{0}\left\{\cos\left[\frac{2\pi}{L}(x_{1}-x_{2})\right]
+2\epsilon\cos\left[\frac{2\pi}{L}(x_{e}-x_{1})\right]
+2\epsilon\cos\left[\frac{2\pi}{L}(x_{e}-x_{2})\right]\right\}.
\end{eqnarray}
We shall insist the the potentials be strongly confining; $V_{0}$ must be
much larger than the kinetic energy scale $\frac{1}{mL^{2}}$. We shall also
require that $\epsilon<1$, but we shall assume that it is
${\cal O}(1)$.

It is convenient to separate out the center of mass motion and define new
coordinates,
\begin{eqnarray}
X & = & \frac{Mx_{1}+Mx_{2}+mx_{e}}{2M+m} \\
y & = & x_{2}-x_{1} \\
x & = & x_{e}-\frac{x_{1}+x_{2}}{2}.
\end{eqnarray}
In these coordinates, we have
\begin{eqnarray}
K & = & -\frac{1}{2(2M+m)}\frac{\partial^{2}}{\partial X^{2}}
-\frac{1}{M}\frac{\partial^{2}}{\partial y^{2}}
-\frac{1}{2\mu}\frac{\partial^{2}}{\partial x^{2}} \\
V & = & V_{0}\left\{\cos\left(\frac{2\pi}{L}y\right)
+2\epsilon\cos\left[\frac{2\pi}{L}\left(x-\frac{y}{2}\right)\right]
+2\epsilon\left[\frac{2\pi}{L}\left(x-\frac{y}{2}\right)\right]\right\}
\nonumber\\
& = & V_{0}\left[\cos\left(\frac{2\pi}{L}y\right)
+4\epsilon\cos\left(\frac{\pi}{L}y\right)\cos\left(\frac{2\pi}{L}x\right)
\right],
\end{eqnarray}
where $\mu\equiv\frac{2mM}{2M+m}$ is a reduced mass for the system.

Writing the full wave function as
\begin{equation}
\Psi(x_{1},x_{2},x_{e})=e^{iPX}\psi(x,y),
\end{equation}
the Hamiltonian governing the relative motion wave function $\psi$ is
\begin{equation}
H_{{\rm rel}}=-\frac{1}{M}\frac{\partial^{2}}{\partial y^{2}}
-\frac{1}{2\mu}\frac{\partial^{2}}{\partial x^{2}}
+V_{0}\left[\cos\left(\frac{2\pi}{L}y\right)
+4\epsilon\cos\left(\frac{\pi}{L}y\right)\cos\left(\frac{2\pi}{L}x\right)
\right].
\end{equation}
Since $\Psi$ is periodic in $x_{1}$, $x_{2}$, and $x_{e}$, $\psi$ must have
the ``helical'' boundary conditions
\begin{eqnarray}
\label{eq-xbound}
\psi(x+L,y) & = & \psi(x,y) \\
\label{eq-ybound}
\psi(x,y+L) & = & \psi(x+L/2,y).
\end{eqnarray}
Although these boundary conditions mix $x$ and $y$, the entire physical
region is contained within the bounds $0\leq x<L$, $0\leq y<L$; so
(\ref{eq-xbound}) and (\ref{eq-ybound}) do not cause any additional mixing of
the $x$ and $y$ dynamics.

We shall analyze the behavior of $\psi$ using the Born-Oppenheimer
approximation. This approximation is justified by the existence of two widely
separated mass scales, $m$ and $M$.
Since $M\gg m\approx\mu$, the atoms move very slowly compared to electron. We
may consider $y$ to be an adiabatically varying parameter and
solve the Schr\"{o}dinger equation governing the $x$ motion.
The Hamiltonian for the electron's relative motion is
\begin{equation}
\label{eq-h''}
H_{{\rm rel}\,e}
=-\frac{1}{2\mu}\frac{d^{2}}{dx^{2}}+4\epsilon V_{0}\cos\left(\frac{\pi}{L}y
\right)\cos\left(\frac{2\pi}{L}x\right).
\end{equation}

We shall distinguish two different parameter regimes. When $\epsilon V_{0}
\left|\cos\left(\frac{\pi}{L}y\right)\right|$ is large compared to $\frac{1}
{\mu L^{2}}$, the potential $V_{{\rm rel}\,e}$
in (\ref{eq-h''}) is strongly confining. However, when $\frac{1}{\mu L^{2}}\gg
\epsilon V_{0}
\left|\cos\left(\frac{\pi}{L}y\right)\right|$, the electron is nearly
free. We must analyze these two cases separately.

We shall assume for now that $y$ lies in the region $0\leq y\leq L/2$. The
minimum value of $\cos\left(\frac{\pi}{L}y\right)$ occurs when $y=
L/2$. In the vicinity of this point, $4\epsilon V_{0}\cos\left(\frac{\pi}{L}
y\right)\approx4\epsilon V_{0}\left(\frac{\pi}{L}\Delta y\right)$, where
$\Delta y\equiv
L/2-y.$ So the transition between the two regimes occurs when
$\Delta y\sim\frac{1}{V_{0}mL}$.  Because $\frac{1}{V_{0}mL^{2}}$ is small,
$\frac{\Delta y}{L}\ll 1$ in both the weakly confined and transition regions.
The potential is strongly confining for all values of $y$, except for in a
comparatively small region around $\Delta y=0$.

Let us first consider the strongly confined regime. The potential
$V_{{\rm rel}\,e}$ has
its minimum at $x=L/2$. Near $x=L/2$, $V_{{\rm rel}\,e}$ has
the form
\begin{equation}
V_{{\rm rel}\,e}(x)\approx4\epsilon V_{0}\cos\left(\frac{\pi}{L}y\right)
\left[-1+\frac{1}{2}\left(
\frac{2\pi}{L}\Delta x\right)^{2}\right],
\end{equation}
for $\Delta x\equiv x-L/2$. So $H_{{\rm rel}\,e}$ becomes
\begin{equation}
H_{{\rm rel}\,e}\approx-\frac{1}{2\mu}\frac{d^{2}}{dx^{2}}-4\epsilon V_{0}\cos
\left(\frac{\pi}{L}y
\right)+\frac{1}{2}\mu\left[\frac{4\pi}{L}\sqrt{\frac{\epsilon V_{0}\cos
\left(\frac{\pi}{L}y\right)}{\mu}}\,\right]^{2}(\Delta x)^{2}.
\end{equation}
This has solutions that are approximately given by harmonic oscillator wave
functions $\phi_{n}(\Delta x)$ centered at $\Delta x=0$. The corresponding
energies are
\begin{equation}
\label{eq-HOenergies}
E_{{\rm rel}\,e,n}\approx-4\epsilon V_{0}\cos\left(\frac{\pi}{L}y\right)+
\left(n
+\frac{1}{2}\right)\left(\frac{4\pi}{L}\right)\sqrt{\frac{\epsilon V_{0}\cos
\left(\frac{\pi}{L}y\right)}{\mu}}.
\end{equation}
The second term in (\ref{eq-HOenergies}) is smaller than the first by a factor
of ${\cal O}\left(\frac{1}{\sqrt{V_{0}mL^{2}}}\right)$, provided
$n+\frac{1}{2}$ is ${\cal O}(1)$.

Now we consider the $\frac{1}{\mu L^{2}}\gg\epsilon V_{0}\left|\cos\left(\frac
{\pi}{L}y
\right)\right|$ regime. In this case, the electron is nearly free. The electron
states are plane waves subject to the boundary condition (\ref{eq-xbound}),
and the energies are
\begin{equation}
E_{{\rm rel}\,e,n}\approx\frac{2\pi^{2}}{\mu L^{2}}n^{2}.
\end{equation}
The ground state corresponds to $n=0$, and the first electronically excited
state is the $n=1$ state that is an odd function of $\Delta x$.

If we extend the energy eigenvalues to cover all possible values of $y$ (not
just $0\leq y\leq L/2$)
and restrict our attention to the two lowest-lying electronic states, we get
\begin{equation}
E_{{\rm rel}\,e,n}\approx\left\{
\begin{array}{l}
-4\epsilon V_{0}\left|\cos\left(\frac{\pi}{L}y\right)\right|+\left(n+\frac{1}
{2}\right)\left(\frac{4\pi}{L}\right)\sqrt{\frac{\epsilon V_{0}\left|\cos\left(
\frac{\pi}{L}y\right)\right|}{\mu}} \\
\frac{2\pi^{2}}{\mu L^{2}}n^{2}
\end{array}
\right.
\end{equation}
The upper expression for $E_{{\rm rel}\,e,n}$ is valid whenever $|y-L/2|
\gg\frac
{1}{\epsilon V_{0}\mu L}$ (taking $0\leq y<L$), while the lower
is correct if
$|y-L/2|\ll\frac{1}{\epsilon V_{0}\mu L}$.
If we include only terms which are zeroth-order in $\frac{1}{V_{0}mL^{2}}$, we
have
\begin{equation}
E_{{\rm rel}\,e,n}\approx-4\epsilon V_{0}\left|\cos\left(\frac{\pi}{L}y
\right)\right|.
\end{equation}
We obtain this simplified form because the two expressions for
$E_{{\rm rel}\,e,n}$ agree near $y=L/2$ in this approximation (i.e. they both
vanish). However, we must still keep in 
mind that this form for $E_{{\rm rel}\,e,n}$ is not quantitatively accurate
near $y=L/2$.

This yields an effective potential for the relative motion of the two atoms,
given by
\begin{eqnarray}
V_{{\rm eff}}(y) & = & V_{0}\left[\cos\left(\frac{2\pi}{L}y\right)-4\epsilon
\left|\cos\left(\frac{\pi}{L}y\right)\right|\right] \nonumber\\
& = & -\left(1+2\epsilon^{2}\right)V_{0}+2V_{0}\left[\left|\cos\left(\frac{\pi}
{L}y\right)\right|-\epsilon\right]^{2}.
\end{eqnarray}
[$V_{{\rm eff}}(y)$ has a cusp at $y=L/2$, but we have already noted that we
expect $E_{{\rm rel}\,e,n}$ to have a slightly
different form in the region around this
point. This
effect should smooth out $V_{{\rm eff}}$ in the vicinity of this local
maximum.]
If $\epsilon<1$, this effective potential has symmetry-breaking minima at
$y_{0}\equiv\frac{L}{\pi}\cos^{-1}\epsilon$ and $L-y_{0}$ (or, equivalently, at
$y_{0}$ and $-y_{0}$). At these points,
$V_{{\rm eff}}(y)=-(1+2\epsilon^{2})V_{0}$. If $\epsilon\geq1$, the
electron-atom repulsion overpowers the atom-atom
repulsion, and the minimum energy configuration has $y=0$; this is possible
because the repulsive cosine potential lacks a hard core region. It was for
this reason that we required that $\epsilon<1$.
We need $\epsilon$ to be ${\cal O}(1)$ so that our formula for
$V'_{{\rm eff}}$
is valid at $y=y_{0}$. That is, $y_{0}$ must lie in the regime where the
electron is tightly confined; if $\epsilon<\frac{1}{\sqrt{V_{0}\mu L^{2}}}$,
then $y_{0}$ will lie too close to $y=L/2$.

The two minima of $y$ correspond to different arrangements of the system.
Since the system is translationally invariant, we may choose to fix $x_{1}=0$,
so that $y$ becomes the position of atom 2. If $y=y_{0}$, the minimum of the
the electron potential $V_{{\rm rel}\,e}$ lies at $x=L/2$. So the electron
will be localized near $x_{e}=L/2+y_{0}/2$. Since $0<y_{0}<L/2$, the order
of the particles as we move in the positive direction is atom 1, atom 2, 
electron; because the system is periodic, this is equivalent to atom 2,
electron, atom 1.
If $y=L-y_{0}$, $V_{{\rm rel}\,e}$ localizes $x$ near $x=0$, so the electron
wave function is peaked at $x_{e}=L/2-y_{0}/2$. The order of the particles
becomes atom 1, electron, atom 2.
In each case, the electron pushes the two atoms
apart. The distance from the atom on the electron's left to the one on the
electron's right (measured through the electron) is $L-y_{0}>L/2$.

Since $V_{{\rm eff}}(y)$ is a even function of $y$, the exact eigenstates of
this potential must be states of definite parity. The parity eigenstates may be
constructed as superpositions
\begin{equation}
\label{eq-superpositions}
\psi(x,y)\approx\frac{1}{\sqrt{2}}[\Phi(y-y_{0})\phi_{n}(x-L/2)\pm\Phi(L-y_{0}
-y)\phi_{n}(x)],
\end{equation}
where $\Phi(y-y_{0})$ is an approximate eigenstate of the Hamiltonian
$H'_{{\rm eff}}=-
\frac{1}{M^{2}}\frac{d^{2}}{dy^{2}}+V'_{{\rm eff}}$, localized around
$y=y_{0}$. These superpositions are not, strictly speaking,
Born-Oppeneheimer states, but they should be good approximations to the
energy eigenstates.
The specific form (\ref{eq-superpositions}) is dependent upon parity
invariance, but the general superposition structure is not. For a general
(asymmetric) potential with two local minima, the energy eigenstates are
superpositions of states localized around those two minima. If $|+\rangle$ is a
state located
at one minimum and $|-\rangle$ is located at the other, then the eigenstates
are $\frac{1}{\sqrt{2}}\left(v_{\pm}|+\rangle\pm v_{\mp}|-\rangle\right)$; the
$v_{\pm}$ are related to the matrix elements of the Hamiltonian by
\begin{equation}
v_{\pm}=\sqrt{1\pm\frac{\delta}{\sqrt{\delta^{2}+4|\langle+|H|-\rangle|^{2}}}},
\end{equation}
where $\delta\equiv\langle+|H|+\rangle-\langle-|H|-\rangle$ is the difference
in energies between the two minima. $\delta$ is assumed to be small compared to
$\langle+|H|+\rangle$ and comparable in
magnitude to $\langle+|H|-\rangle$.

The form (\ref{eq-superpositions}) for $\psi$ holds in the
region $0\leq y\leq L$, $0\leq x\leq L$; it may be extended to other
values of $x$ and $y$ using the boundary
conditions~(\ref{eq-xbound}--\ref{eq-ybound}). There are no low-lying energy
eigenstates for which the mean atomic separation is $L/2$ that do not have
this superposition form, because $y=L/2$ is a local maximum of
$V'_{{\rm eff}}$.

The two superposition states in
(\ref{eq-superpositions}) are not degenerate. There will be some
mixing between the two states, shifting the energy of each. We shall now show
that this energy shift is very small, so that the expressions given in
(\ref{eq-superpositions}) are very good approximations to the exact wave
functions.

In the WKB approximation, the energy difference between the two states is
$\Delta E=
\frac{\omega T}{\pi}$~\cite{ref-landau}; $\omega$ is the frequency of
classical oscillations about the minima, and $T$ is the tunneling amplitude
$T=\exp\left[-\int\sqrt{M(V_{{\rm eff}}-E)}\, dx\right]$, where the integration
extends over the barrier region. The small oscillation frequency $\omega$ is
just
\begin{eqnarray}
\omega & = & \sqrt{\left(\frac{2}{M}\right)\left.\frac{d^{2}}{dy^{2}}V_{{\rm
eff}}(y)\right|_{y=y_{0}}} \nonumber\\
& = & \left(\frac{2\pi}{L}\right)\sqrt{\frac{2V_{0}(1-\epsilon^{2})}{M}}.
\end{eqnarray}
Because of the model's periodic boundary conditions, $T$ is actually composed
of two terms, $T_{1}$ and $T_{2}$. These correspond to the amplitude for the
particle to tunnel to the right (from $y_{0}$ to $L-y_{0}$) and for it to
tunnel to the left (from $y_{0}$ to $-y_{0}$), respectively. If 
we neglect the quantum fluctuations around the minima, then
$T$ is given by
\begin{eqnarray}
T & = & T_{1} + T_{2} \nonumber\\
T & = & \exp\left\{-\int_{y_{0}}^{L-y_{0}}dy\,\sqrt{M[V_{{\rm eff}}(y)-V_{{\rm
eff}}(y_{0})]}\right\}+\exp\left\{-\int_{-y_{0}}^{y_{0}}dy\,\sqrt{M[V_{{\rm
eff}}(y)-V_{{\rm eff}}(y_{0})]}\right\} \nonumber\\
& = & \exp\left[-\frac{2L\sqrt{2MV_{0}}}{\pi}\int_{\cos^{-1}\epsilon}^{\frac
{\pi}{2}}du\,(\epsilon-\cos u)\right]+
\exp\left[-\frac{2L\sqrt{2MV_{0}}}{\pi}\int_{0}^{\cos^{-1}\epsilon}du\,
(\cos u-\epsilon)\right] \nonumber\\
& = & \exp\left[-\frac{2L\sqrt{2MV_{0}}}{\pi}\left(\epsilon\sin^{-1}\epsilon+
\sqrt{1-\epsilon^{2}}-1\right)\right] \nonumber\\
& & +\exp\left[-\frac{2L\sqrt{2MV_{0}}}{\pi}\left(\sqrt{1-
\epsilon^{2}}-\epsilon\cos^{-1}\epsilon\right)\right].
\end{eqnarray}
We conclude that the energy difference between the two states, $\Delta E=\frac
{\omega(T_{1}+T_{2})}{\pi}$ is an exponentially small function of the large
parameter $\sqrt{V_{0}ML^{2}}$, so the mixing between the two superposition
states is minimal.

We may now relate our toy model to the liquid helium problem. In each system,
the electron repels the atoms surrounding it.
The electron-filled bubble in the liquid helium system corresponds to the
``bubble'' the
electron in the toy model creates by forcing apart the atoms on its left and
its right.

The states corresponding to (\ref{eq-superpositions}) in the liquid helium
system are superpositions of different position states of the bubble. A single
bubble exists at either of two locations; the two spatially separated bubbles
do not coexist simultaneously. A classical treatment of the bubble fails to
account for these
states. The analogue in the toy model of treating the bubble as a purely
classical object is
the assumption that the wave function must be an unentangled product of a
function of $x$ and a function of $y$. Specifically, the two-bubble,
split-electron state from~\cite{ref-maris} corresponds to the toy model state
\begin{equation}
\psi_{{\rm class}}(x,y)=\frac{1}{2}[\Phi(y-y_{0})+\Phi(L-y_{0}-y)][\phi_{0}(x)
-\phi_{0}(x-L/2)],
\end{equation}
which is clearly not an eigenstate of the energy.

The analogue of $\psi_{{\rm class}}(x,y)$ for the helium system may be written
schematically as
\begin{equation}
\label{eq-psiHeM}
\psi^{{\rm He}}_{{\rm class}}({\bf X}_{b},{\bf X}_{e})=\frac{1}{2}[\psi^{b}_{1}
({\bf X}_{b})+\psi^{b}_{2}({\bf X}_{b})][\psi^{e}_{1}({\bf X}_{e})-\psi^{e}_{2}
({\bf X}_{e})].
\end{equation}
${\bf X}_{b}$ and ${\bf X}_{e}$ are the coordinates of the bubble and electron,
respectively, while $\psi^{b}_{i}$ and $\psi^{e}_{i}$ are appropriate wave
functions, localized at two
different positions indexed by $i$. It is easy to see that this wave function
is not an energy eigenstate. The $\psi^{b}_{1}({\bf X}_{b})\psi^{e}_{2}({\bf X}
_{e})$ and $\psi^{b}_{2}({\bf X}_{b})\psi^{e}_{1}({\bf X}_{e})$ terms
correspond to the
electron and the bubble being in different locations.  These configurations
are unstable; the empty bubble will rapidly collapse, and a new bubble will
form around the electron. This process will occur even while the two bubbles
are still splitting apart; as the bubble splits, the bubble and electron wave
functions will become entangled. For this reason, it is incorrect to treat the
bubble classically.

We also note that, if (\ref{eq-psiHeM}) were a stationary state, it would
be possible to send faster-than-light signals between the two bubbles. We would
begin by separating the two bubbles by a large distance. Then we measure
whether or not the electron is present within one bubble. If we find the
electron in that bubble, we know that the electron is not present in the other
bubble, and the other bubble consequently will collapse.  Similarly, if we do
not find the electron in the first bubble, it must be in the second bubble.
The second bubble with then expand, because it now contains an entire electron.
In either case, our measurement has affected the size of the second bubble in
a measurable way, so someone observing the second bubble would immediately know
that we had performed the measurement\footnote{We would like to
thank S. Glashow for pointing out the acausal nature of this situation}. Of
course, this paradox does not arise
for the superposition wave functions
\begin{equation}
\label{eq-psiHe}
\psi^{{\rm He}}({\bf X}_{b},{\bf X}_{e})=\frac{1}{\sqrt{2}}[\psi^{b}_{1}
({\bf X}_{b})\psi^{e}_{1}({\bf X}_{e})\pm\psi^{b}_{2}({\bf X}_{b})\psi^{e}_{2}
({\bf X}_{e})];
\end{equation}
a measurement of the electron's position reveals a full-sized bubble at the
location where the electron is found and no bubble at all in the other
location.

Finally, we point out that the superposition states displayed in
(\ref{eq-superpositions}) are exactly the sort of
states discussed in~\cite{ref-jackiw2}.
If we attempt to measure the fermion number
between atom 1 on the left and atom 2 on the right, we find that the
expectation value is fractional---with value $\frac{1}{2}$. However,
the presence of the two nearly degenerate
superposition states leads to a large dispersion in this localized fermion
number. An analogous phenomenon occurs for the liquid helium wave function
(\ref{eq-psiHe}). In each case, the large dispersion indicates that the
fractionization is a characteristic of the
expectation values only---not of the eigenvalues.

\section*{Acknowledgments}

We wish to thank R. Jackiw for many helpful discussions.
This work is supported in part by funds provided by the U. S. Department of
Energy (D.O.E.) under cooperative research agreements DE-FC02-94ER40818
and DE-FG02-91ER40676.

\end{document}